\documentstyle[sprocl,epsfig]{article}

\def\Journal#1#2#3#4{{#1} {\bf #2}, #3 (#4)}

\def\NPB{{\em Nucl. Phys.} B}

\def\PRL{\em Phys. Rev. Lett.}
\def\PRD{{\em Phys. Rev.} D}
\def\ZPC{{\em Z. Phys.} C}

\def\bsgam{b\rightarrow s\gamma}
\def\as{\alpha_s}
\def\be{\begin{equation}}
\def\ee{\end{equation}}
\def\bea{\begin{eqnarray}}
\def\eea{\end{eqnarray}}

\def\sign{{\sf sign}}
\def\uln{\underline}
\def\co#1{{\ifmmode{\cal O}_{#1}\else${\cal O}_{#1}$\fi}}

\begin{document}

\begin{titlepage}
\mbox{ }
\vspace{-2.0 cm}

\begin{tabbing}
xxxxxxxxxxxxxxxxxxxxxxxxxxxxxxxxxxxxxxxxxxxxxxxxxxxx\=xxxxxxxxxx\kill
 \> IUHET-375 \\ 
 \> OHSTPY-HEP-T-97-022 \\ 
 \> \hfill November 1997 \\
\end{tabbing}


\vspace{1.5 cm}

\begin{center}

\renewcommand{\thefootnote}{\fnsymbol{footnote}}

{\Large\bf Global Analysis of Electroweak Data in SUSY GUTs}~\footnote
{Talk presented by Tom\'{a}\v{s} Bla\v{z}ek at the International 
 Workshop on Quantum Effects in the MSSM, Barcelona, Catalonia, Spain,
 9-13 September 1997. To appear in the proceedings.}

\vspace{1.0 cm}

{\large\bf  Tom\'{a}\v{s} Bla\v{z}ek~\footnote
            {On leave of absence from the Dept.\ of Theoretical Physics, 
            Comenius Univ., Bratislava, Slovakia; current e-mail address: 
            {\em blazek@gluon2.physics.indiana.edu}}}\\ 
{\em Department of Physics, Indiana University, Swain Hall West 117, Bloomington, IN 47405, USA}
\vskip 0.2cm
{\large\bf  and} 
\vskip 0.2cm
{\large\bf  Stuart Raby~\footnote
            {current e-mail address: 
            {\em raby@pacific.mps.ohio-state.edu}}}\\ 
{\em Department of Physics, The Ohio State University, 174 W. 18th Ave., Columbus, OH 43210, USA}


\vspace{1.0 cm}


{\bf Abstract}

\end{center}

A $\chi^2$ analysis of several SUSY GUTs recently discussed
in the literature is presented. We obtain global fits to electroweak data,
which include gauge couplings, gauge boson masses, $BR(\bsgam)$ and masses 
of fermions of all three generations and their mixing angles. Thus 
we are able to test gauge unification, radiative electroweak symmetry 
breaking, SUSY sector ( -- in the context of supergravity induced SUSY
breaking)  and the Yukawa sector in each particular model
self-consistently. One of the models studied provides a very good fit
with $\chi^2\sim 1$ for $3$ degrees of freedom, in a large
region of the allowed SUSY parameter space. The Yukawa sector works so well
in this case that the analysis ends up testing the MSSM
constrained by unification. Adopting this point of view,
in the second part of this talk we focus on the details of the fit 
for $BR(\bsgam)$ and discuss the correlations among $\delta m_b^{SUSY}$, 
$\alpha_s(M_Z)$ and a GUT threshold to $\alpha_s(M_G)$.
We conclude that an attractive $SO(10)$-derived regime of the MSSM
remains a viable option.

\vfill

\setcounter{footnote}{0}
\renewcommand{\thefootnote}{\arabic{footnote}}

\end{titlepage}

%
%

\title{
GLOBAL ANALYSIS OF ELECTROWEAK DATA IN SUSY GUTS
}

\author{T. BLA\v{Z}EK}

\address{Department of Physics, Indiana University,\\
         Swain Hall West 117, Bloomington, IN 47405\\
         E-mail: blazek@gluon2.physics.indiana.edu} 

\author{S. RABY}

\address{Department of Physics, The Ohio State University,\\
         174 W. 18th Ave., Columbus, OH 43210 \\
         E-mail: raby@pacific.mps.ohio-state.edu}

\maketitle\abstracts{
A $\chi^2$ analysis of several SUSY GUTs recently discussed
in the literature is presented. We obtain global fits to electroweak data,
which include gauge couplings, gauge boson masses, $BR(\bsgam)$ and masses 
of fermions of all three generations and their mixing angles. Thus 
we are able to test gauge unification, radiative electroweak symmetry 
breaking, SUSY sector ( -- in the context of supergravity induced SUSY
breaking)  and the Yukawa sector in each particular model
self-consistently. One of the models studied provides a very good fit
with $\chi^2\sim 1$ for $3$ degrees of freedom, in a large
region of the allowed SUSY parameter space. The Yukawa sector works so well
in this case that the analysis ends up testing the MSSM
constrained by unification. Adopting this point of view,
in the second part of this talk we focus on the details of the fit 
for $BR(\bsgam)$ and discuss the correlations among $\delta m_b^{SUSY}$, 
$\alpha_s(M_Z)$ and a GUT threshold to $\alpha_s(M_G)$.
We conclude that an attractive $SO(10)$-derived regime of the MSSM
remains a viable option.
}

\section{Introduction}

Reaching beyond the Minimal Supersymmetric Standard 
Model (MSSM) we apply global analysis to the search for 
grand unified theories (GUTs). That enables us to compare 
specific GUT models which become the MSSM as effective theory below 
the unification scale. In practice, it means that we test gauge 
unification, radiative electroweak symmetry breaking and the SUSY sector in 
the same way as the MSSM analysis constrained by unification \cite{gaMSSM}, 
but in addition 
we also test the Yukawa sector of the theory versus the observed fermion
masses of all three generations and the Cabibbo-Kobayashi-Maskawa (CKM)
matrix elements. Clearly, we agree with the MSSM analysis for
models which describe the Yukawa sector very well. 

There are plenty of models to be tested in this kind of analysis.
For the beginning we start with models based on SO(10) gauge symmetry. 
SO(10) SUSY GUTs have been recognized as excellent candidates for an effective 
field theory below the Planck/string scale. They maintain the successful
prediction for gauge coupling unification and provide a powerful and very
economic framework for theories of fermion masses. This is because 
all the fermions of one generation are contained in the 16 dimensional
representation of SO(10) - thus fermion mass matrices are
related by symmetry. In the most predictive theories,
the ratio of Higgs vevs - tan$\beta$ - is large, and the
top quark is naturally heavy as found experimentally. However, 
as a consequence of large tan$\beta$ there are potentially large
supersymmetric one-loop effects at the weak scale which could
 play an important role in fitting the fermion masses and
mixings, and the FCNC processes like the observed $\bsgam$ decay rate. 
Thus a self-consistent analysis becomes more powerful
(and restrictive) than in a low tan$\beta$ scenario.
 
In sections 3 and 4, we present the results of such a 
complete top-down analysis. In section 5, we focus on the MSSM 
constrained by the best working GUT model and analyze it similarly 
to generic MSSM analyses constrained by unification. 
\footnote
{
However, our MSSM analysis is simplified. 
We give up on most of the precision
electroweak observables and keep
only ten observables of the MSSM
analysis (see the next section for
details). We believe that this 
reduction does not bias our 
results in any significant way since
the asymmetries and $Z$ lineshape 
parameters do not present dominant
constraints, especially if most of 
the SUSY particles are rather heavy
(as happens for the best working 
model of our analysis).
}
We present the best fit correlations among 
various contributions to the $\bsgam$ decay rate, SUSY corrections to 
$m_b$, and rather low values of $\as(M_Z)$ -- all in the $(m_0,M_{1/2})$
SUSY parameter space. The phenomenological implications are discussed 
as well.

\section{Global Analysis}

Details of our numerical procedure are described in \cite{gaGUT}. The 
analysis starts at the GUT scale $M_G$, which is a free parameter itself, 
with unified gauge coupling $\alpha_G$, $n_y$ free parameters entering 
the Yukawa matrices\footnote
{
Clearly, $n_y$ is 
model dependent.
}, 
and with $\epsilon_3$ as a one loop GUT threshold correction to 
$\alpha_s(M_G)$ \footnote
{
$M_G$ is defined as the scale 
where the gauge couplings 
$\alpha_1$ and $\alpha_2$ 
are exactly equal within the 
one-loop GUT threshold 
corrections. By $\alpha_G$ 
we actually mean the value 
$\alpha_1(M_G)\equiv 
            \alpha_2(M_G)$
}.
We assume supergravity induced SUSY breaking. At the scale $M_G$ we 
introduce standard universal soft SUSY breaking parameters $m_0,\: M_{1/2}$ 
and $A_0$, and non-universal Higgs masses $m_{H_d}$ and $m_{H_u}$.
The $\mu$ parameter and its SUSY breaking bilinear partner
$B$ are introduced at the $Z$ scale, since they are renormalized 
multiplicatively and do not enter the RGEs of the other parameters.
At the $Z$ scale we match the MSSM directly to $SU(3)_c\times U(1)_{em}$, 
thus leaving out the SM as an effective theory$\,$\cite{cpp}. 
Electroweak symmetry breaking 
is established at one loop in the process of the $\chi^2$ minimization by 
fitting the observables in table \ref{t_obs}. Within 
the MSSM, we calculate one-loop corrected $W$ and $Z$ masses and $G_{\mu}$, 
corrections to the $\rho$ parameter from new physics outside the SM, 
and the amplitude for the process $\bsgam$.
When crossing the $Z$ scale, we compute 
the complete one loop threshold corrections to  $\alpha_s$ and 
$\alpha$, whereas only those one loop threshold corrections to the 
fermion masses and mixings enhanced by tan$\beta$ are computed$\,$\cite{bpr}. 
\begin{table}[t]
\caption{Experimental observables of the global analysis.\label{t_obs}}
$$ 
\begin{array}{|r|c|c|c||r|c|c|c|}
\hline
\multicolumn{2}{|c|}{\rm Observable} & {\rm Central}& \sigma &
\multicolumn{2}{ c|}{\rm Observable} & {\rm Central}& \sigma \\
\multicolumn{2}{|c|}{  }             & 
\makebox[1.7cm]{\rm value}  & \makebox[1.7cm]{} &
\multicolumn{2}{ c|}{  }             & 
\makebox[1.7cm]{\rm value}  & \makebox[1.7cm]{} \\
\hline
 1. & M_Z              &  91.186      & \uln{0.46}      &
11. & M_b - M_c       &    3.4        &  0.2       \\
 2. & M_W              &  80.356      & \uln{0.40}      &
12. & m_s             &  180          & 50         \\
 3. & G_{\mu}    &  1.166\cdot 10^{-5} &\uln{1.2\cdot 10^{-7}} &
13. & m_d/m_s         &  0.05         &  0.015     \\
 4. & \alpha^{-1}     &  137.04       & \uln{0.69}      &
14. & Q^{-2}          &  0.00203      &  0.00020   \\
 5. & \alpha_s(M_Z)   &  0.118        &  0.005     &
15. & M_{\mu}         & 105.66        & \uln{0.53}      \\
\cline{1-4}
 6. & M_t             &  175.0        &  6.0       &
16. & M_e             &  0.5110       & \uln{0.0026}    \\
\cline{5-8}
 7. & m_b(M_b)        &    4.26       &  0.11      &
17. & V_{us}          &  0.2205       &  0.0026    \\
 8. & M_{\tau}        &  1.777        & \uln{0.0089}    &
18. & V_{cb}          &  0.0392       &  0.003     \\
\cline{1-4}
 9. & \rho_{new} & -0.6 \cdot 10^{-3}  & 2.6\cdot 10^{-3}   &
19. & V_{ub}/V_{cb}   &  0.08         &  0.02      \\
10. & B(b \rightarrow s \gamma) &  2.32\cdot 10^{-4} &  0.92\cdot 10^{-4}  &
20. & \hat B_K        &  0.8          &  0.1       \\
\hline
\end{array}
$$
\end{table}
The $Z$-scale amplitude for $\bsgam$ is matched to the 
coefficient $C_7(M_Z)$ which is then 
renormalized down to the scale $M_b$,
\be
  C_7^{eff}(M_b) = \eta^{16\over23}C_7(M_Z) 
                  +{8\over 3}\:(\eta^{14\over23} - \eta^{16\over23})\:C_8(M_Z)
                  +C_2(M_Z)\,\sum_{i=1}^8h_i\eta^{a_i}\,,
\label{C7eff}
\ee
based on mixing of the electromagnetic operator with the chromomagnetic 
and current-current operators (coefficients $C_8$ and $C_2$) in the 
leading log approximation; with $\eta= \as(M_Z)/\as(M_b)$. 
Finally, the branching ratio
\be
 BR(\bsgam) = \frac{|V^*_{ts}V_{tb}|^2}{|V_{cb}|^2}\;
              \frac{6\alpha}{\pi g(M_c/M_b)}\; |C_7^{eff}(\mu_b)|^2\;
              BR(b\rightarrow ce\bar{\nu})\;,
\label{BR}
\ee
where $BR(b\rightarrow ce\bar{\nu})=0.104$, $\alpha=1/132.5$, 
the phase-space function $g(z)=1-8z^2+8z^6-z^8-24z^4\log z$, and  
$h_i$'s and $a_i$'s in eq.(\ref{C7eff}) are given in ref.\cite{buras}. 
The values of the CKM matrix elements and quark masses in (\ref{BR}) are 
consistently calculated in the actual fit within a particular model. 

Note that $M_Z,\: M_W,\: \alpha,\: G_{\mu}$ and the lepton masses are known 
so well that we have to assign a theoretical error as their standard deviation
( --- the corresponding $\sigma$'s are underlined in table \ref{t_obs}). 
We estimated 
conservatively the theoretical error to be 0.5\% based on the uncertainties 
from higher order perturbation theory and from the performance of our 
numerical analysis. The error on $G_{\mu}$ is estimated to be within 1\% 
due to the fact that in addition to the uncertainties mentioned above 
we neglect SUSY vertex and box corrections to $\Delta r$.
Also note that $\epsilon_K$, the observable of $CP$ violation, has been 
replaced by a less precisely known hadronic matrix element $\hat{B}_K$ . 
Similarly, the light quark masses are replaced by 
their ratios (at the scale 1GeV),  and $m_c(M_c)$ by the difference 
$M_b-M_c$, since the latter quantities are known to better accuracy. 
$Q^{-2}\doteq (m_d^2-m_u^2)/m_s^2$ is the Kaplan-Manohar-Leutwyler ellipse 
parameter. Finally note, that the data in table \ref{t_obs} 
are divided into two groups. The first ten, {\em i.e.}
five observables in the gauge sector ($M_Z,\,M_W,\,G_\mu,\,\alpha$ and $\as$), 
masses of the third generation fermions, $\rho_{new}$ and $BR(\bsgam)$ 
can also be included in tests of the MSSM constrained by unification 
with no particular underlying GUT. 
The other ten correspond to six light fermion masses and four 
independent parameters of the CKM matrix and test
primarily the Yukawa sector of a GUT model.

In addition, the $\chi^2$ function is increased significantly by a special 
penalty whenever a sparticle mass goes below its experimental limit.

\section{Results for SO(10) GUT Models with Four
         Effective Operators} 

The analysis, as described above, was used to test simple SO(10) models.
First we checked the performance of the nine models with four effective
operators suggested in \cite{adhrs}. The models were defined by a unique
choice of the operators $\co{33}= 16_3\:10\:16_3$ and 
$\co{12}= 16_1\:(\frac{\tilde{A}}{S})^3\:10\: (\frac{\tilde{A}}{S})^3\:16_2$
at the GUT scale ($A$'s stand for adjoint states and $S$ for singlets).
There were six \co{22} operators which all gave the same 0:1:3 
Clebsch relation between the $22$ elements of the Yukawa matrices 
for the up quarks, down quarks and charged leptons, thus introducing
Georgi-Jarlskog relation in all models. Finally, the models were 
distinguished by a choice of the \co{23} operator. Nine 
operators were suggested in ref.\cite{adhrs} leading to
different predictions at low energies. 

Our analysis has shown that the choice of the \co{23} operator is indeed 
significant (see fig.1a) and that model 4 is by far the best working model. 
In this case, 
$\co{23} = 16_2\:(\frac{A_2}{\tilde{A}})\:10\: (\frac{A_1}{\tilde{A}})\:16_3$.
Note that with four effective operators we have $n_y=5$ free parameters
in the Yukawa matrices at the GUT scale. That leaves the $\chi^2$ function
(out of the 20 observables in table \ref{t_obs}) with 5 degrees 
of freedom (d.o.f.).
We show in fig.1b that the performance of model 4 does not get 
significantly better in a larger SUSY parameter space. We checked that 
the same is true for the other models. We also checked that no substantial
improvement of the performance of model 4 can be achieved by neglecting one 
out of the twenty observables given in table \ref{t_obs} \cite{gaGUT}. 
On the other hand, we have found 
that a significant improvement is possible by adding one new operator,
contributing to the 13 and 31 elements of the Yukawa matrices.

\section{Results for SO(10) GUT Models with Five
         Effective Operators} 

Next, we analyzed two models recently derived from the complete SO(10) 
SUSY GUTs \cite{lucas}. The models were constructed 
\protect
\begin{figure}[tb]
\centering
\mbox{%
\epsfig{file=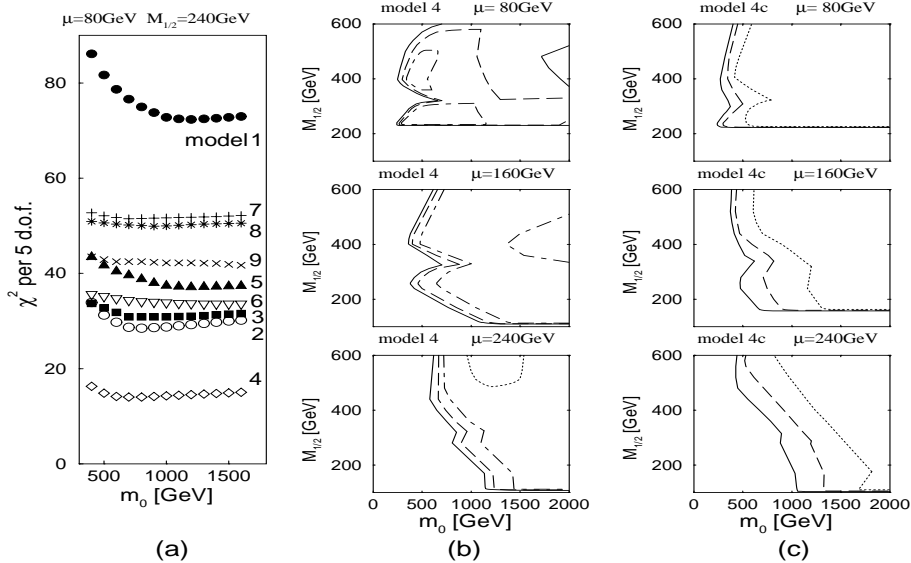 
        ,height=7cm  
        ,width=11cm   
       }%
}
\caption{Results of global analysis for SO(10) GUT models: 
(a) nine models of section 3;
(b) model 4 [solid (dashed, short-long-dashed, dotted) lines
represent contour lines of constant $\chi^2=16\,(15,\,14,\,13)$ per 5 d.o.f.];
(c) model 4c [solid (dashed, dotted) lines
represent contour lines of constant $\chi^2=6\,(3,\,1)$ per 3 d.o.f.].} 
\label{fig1}
\end{figure}
as simple extensions of model 4. Different label 
\mbox{(a,b,...f)} refers to the different possible 22 operators. In
the extension to a complete GUT the different 22 operators lead to 
inequivalent theories due to different U(1) charge assignments. 
When one demands ``naturalness'', i.e. includes all terms in the superpotential
 consistent with the symmetries of the theory one finds one and only one 
13 operator for models 4a and 4c. The 22 and 13 operators of
model 4c are 
\mbox{$ \co{22}=16_2\:{\tilde A \over S}\:10\:{A_1\over S}\:16_2$} and 
\mbox{$ \co{13}=16_1\:({\tilde A \over S})^3\:10\:{A_2 \over S}\:16_3\;\;$}. 
With the 13 operator $n_y$=7, which implies 3 degrees of freedom. The 
results of the global analysis are given in fig.1c. 
Similarly to fig.1b, these figures show the contour plots of 
the minimum $\chi^2$ in the $(m_0,M_{1/2})$ plane, for three different 
fixed values of the parameter $\mu(M_Z)=80,\,160,\, 240$ GeV.
All initial parameters other than \{$m_0,M_{1/2},\mu$\} were subject 
to minimization. 

The fits get worse as $\mu$ increases because the SUSY corrections to 
fermion masses and mixings increase with $\mu$. As
$\mu$ gets larger they can only be kept under control by larger squark masses. 
Varying $\mu$ freely actually results in its approaching 
the lowest possible value. The lower bound on $\mu(M_Z)$ is determined by 
the chargino mass limit from direct searches and is correlated with $M_{1/2}$. 
When the value of $\mu$ is fixed, as in fig's 1b-c, the chargino mass 
limit then sets a sharp lower bound on $M_{1/2}$, which is explicitly 
visible in each of the figures. Figures 1b and 1c were obtained assuming 
$m_{\chi^-} > 65$GeV. In our further analysis $\mu(M_Z)$ has been fixed 
to $110$GeV and $m_{\chi^-} > 85$GeV was imposed.
Figures 2a-b show explicitly that the structure observed in fig.1c 
originates from the two distinct fits corresponding to two separate minima
of the global analysis. The fits are primarily distinguished by the sign
of the $\bsgam$ decay amplitude. ( See section 5 for more details.) 
We do not show the results for negative values of $\mu$. In this case,
the chargino contribution to $\bsgam$ interferes constructively with 
the already large enough SM and charged Higgs contributions. As a result, 
the fits get much worse, with $\chi^2$ well above 
10 per 3 d.o.f., and that disfavors this region of the SUSY parameter space.
Similar observations were also made by W.~Hollik {\em et al} \cite{gaMSSM}.
 
Model 4a is defined by different 22 and 13 operators and gives the fits 
with the best $\chi^2\simeq$4-6 in most of the parameter space. 
(It yields also $\chi^2\simeq$3, but only for the corner in the SUSY 
parameter space with large $m_0,\: M_{1/2}$ and $\mu$.)$\,$\cite{gaGUT}

Whether or not these particular models are close to the path Nature has 
chosen remains to be seen. One important test will be via the CP violating 
decays of the B meson. Models 4c and 4a both predict a narrowly spread 
value sin2$\alpha\simeq0.95$, whereas in the 
SM the value of sin2$\alpha$ is unrestricted$\,$\cite{ali}. Another important 
test may come from nucleon decay rates \cite{lr2}.

\section{MSSM Analysis Constrained by the Best Working GUT Model}

Since the fermionic sector of model 4c works very well, 
we can regard the analysis in this case as a test of the MSSM with large
tan$\beta$. Yet, some model dependence is still present. 
It is, first of all, the
introduction of $\epsilon_3$, a GUT threshold to $\as$. 
It is actually the only GUT threshold introduced in our study. 
Note in subsection \ref{ss_e3} that non-zero $\epsilon_3$ is imposed 
by the low energy phenomenology rather than by physics at the GUT scale. 

For the Yukawa matrices, the exact equality of the 33 elements is assumed.
The remaining Yukawa entries are
small and decouple from the MSSM RGEs for the gauge and third generation 
Yukawa couplings and diagonal SUSY mass parameters. 
Thus they have no effect on the calculation of the $Z$-scale values 
for the first nine observables in table \ref{t_obs}. 
The $BR(\bsgam)$ is affected by some of these entries. 
Model dependence comes from the chargino contribution which contains
inter-generational $\tilde{c}_L$-$\tilde{t}_L$ squark 
mixing. That contribution is tan$\beta$ enhanced which makes it non-negligible.
The mixing is completely induced by the off-diagonal 
entries of the Yukawa matrices in the RG evolution.
Next, note that the light quark and lepton masses and CKM elements do not exert
any significant pull on the best $\chi^2$'s of model 4c \cite{gaGUT}.
We conclude 
that our results presented in this section are not sensitive to the
structure of the Yukawa matrices except for the model
dependent 23 mixing which is significant for the $BR(\bsgam)$.

\subsection{Results for $BR({\bsgam})$}
We find that
the $\bsgam$ amplitude is dominated by the SM, 
charged Higgs, and tan$\beta$ enhanced chargino contributions.
To understand our results, let's estimate first what to expect from 
the SUSY contribution to $C_7(M_Z)$ and define the $Z$ scale ratios
for Higgs and chargino contributions $r^{(H)} = C_7^{(H)} / C_7^{(SM)}$ and 
$r^{(C)} = C_7^{(C)} / C_7^{(SM)}$.
Equation (\ref{C7eff}) then reads
\be
  C_7^{(MSSM)\:eff} \approx \eta^{16\over23}\:C_7^{(SM)}\,
                                       (1+ r^{(H)} + r^{(C)}) 
                      +\sum_{i=1}^8h_i\eta^{a_i}\,
\label{C7eff_MSSM}
\ee
where we used $C_2(M_Z)=1$ and $C_8\ll C_7,\,C_2$. Since it is well known 
that $C_7^{(SM)\:eff}\:(\approx 
\eta^{16\over23}\,C_7^{(SM)} +\sum_{i=1}^8h_i\eta^{a_i})$ 
would yield about the right value for the $BR(\bsgam)$ we infer that either
\bea
   r^{(C)} &\approx& - r^{(H)}\,,\:\:\:\:\:\:\:\:\:\:\:\:\:
                                        \:\:\: \mbox{\rm for}\: 
   C_7^{(MSSM)\:eff} \approx +\, C_7^{(SM)\:eff}\,,
\label{r_C_ac} 
\\
\mbox{\rm or  }\;\;\;\;\;\;\;\;\;\;\;\;\;\;&\mbox{ }&\;
\nonumber
\\ 
   r^{(C)} &\approx& - r^{(H)} - 4.60\,,\:\:\: \mbox{\rm for}\: 
   C_7^{(MSSM)\:eff} \approx - \, C_7^{(SM)\:eff}.
\label{r_C_AC}
\eea
For the last 
estimate, the numerical results $C_7^{(SM)}=-0.190$, $\eta^{16\over23}=0.679$,
and  $\sum_{i=1}^8h_i\eta^{a_i}\,= -0.168$ were used --- computed for 
$\as(M_Z)=0.118$. 
 
The charged Higgs contribution always interferes constructively with the SM 
contribution. We get 
$0 < r^{(H)} < 1.3$, 
depending on the mass of the $H^-$. In the first case, especially  
if $r^{(H)}$ and $r^{(C)}$ are non-negligible, eq.(\ref{r_C_ac})
means that the chargino part must interfere destructively\footnote
{Since eq's (\ref{C7eff_MSSM}) 
and (\ref{r_C_ac}) are valid only 
approximately, the case 
$1+ r^{(H)} + r^{(C)}\approx 1$
in principle also allows for a 
constructive interference 
$0 < r^{(C)},\: r^{(H)} \ll 1$
in the region in parameter space 
where $m_{H^-}$ and sparticle masses
are large. This option, however, 
does not result from our best fits, 
as already mentioned in the 
discussion on negative $\mu$ 
parameter in the previous section. 
}
with the SM and charged Higgs contributions,
practically cancelling the latter. The enhancement by tan$\beta$ of the 
chargino contribution has to be compensated for by rather large sparticle 
masses. In the second case, described by eq.(\ref{r_C_AC}), large destructive
chargino interference is required to outweigh the combined SM and $H^-$
contributions and to flip over the overall sign of the amplitude. 
Quite amazingly, it is not so difficult to arrange 
since the chargino contribution is the only one enhanced by
large tan$\beta$. However, large sparticle masses obviously suppress
the effect. The lesson is that we can expect the two cases to work in
a complementary SUSY parameter space and have 
\sign$\,C_7^{(MSSM)} = 
    \stackrel{+}{\scriptstyle (-)}$ \sign$\,C_7^{(SM)}$ 
for the best fits in the region with large (low) $m_0$ and/or $M_{1/2}$, 
respectively.
\protect
\begin{figure}[p]
\centering
\mbox{%
\epsfig{file=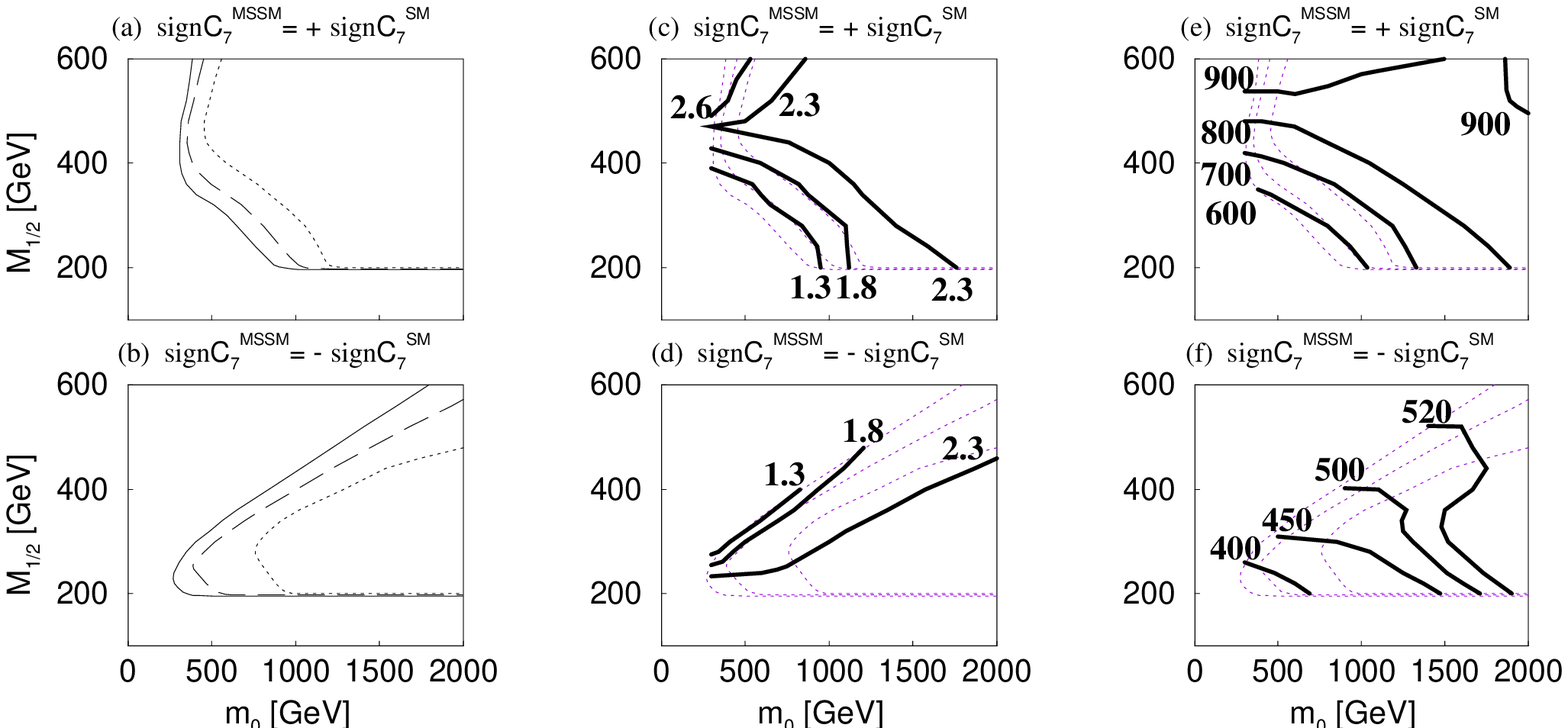 
        ,height=3.5cm  
        ,width=10cm   
       }%
}
\caption{Model 4c global analysis results 
for the two possible signs of $C_7^{MSSM}$.
The best fit contour plots of
(a),(b) $\chi^2$ 
        [solid (dashed, dotted) lines correspond to
        $\chi^2=6\,(3,\,1)$ per 3 d.o.f.];
(c),(d) BR$(\bsgam)\times 10^4$ ;
(e),(f) the lightest stop mass $m_{\tilde{t}_1}$ [in GeV]; 
        with the $\chi^2$ contour lines in the background.}
\label{fig2}
\vskip 11mm
\mbox{%
\epsfig{file=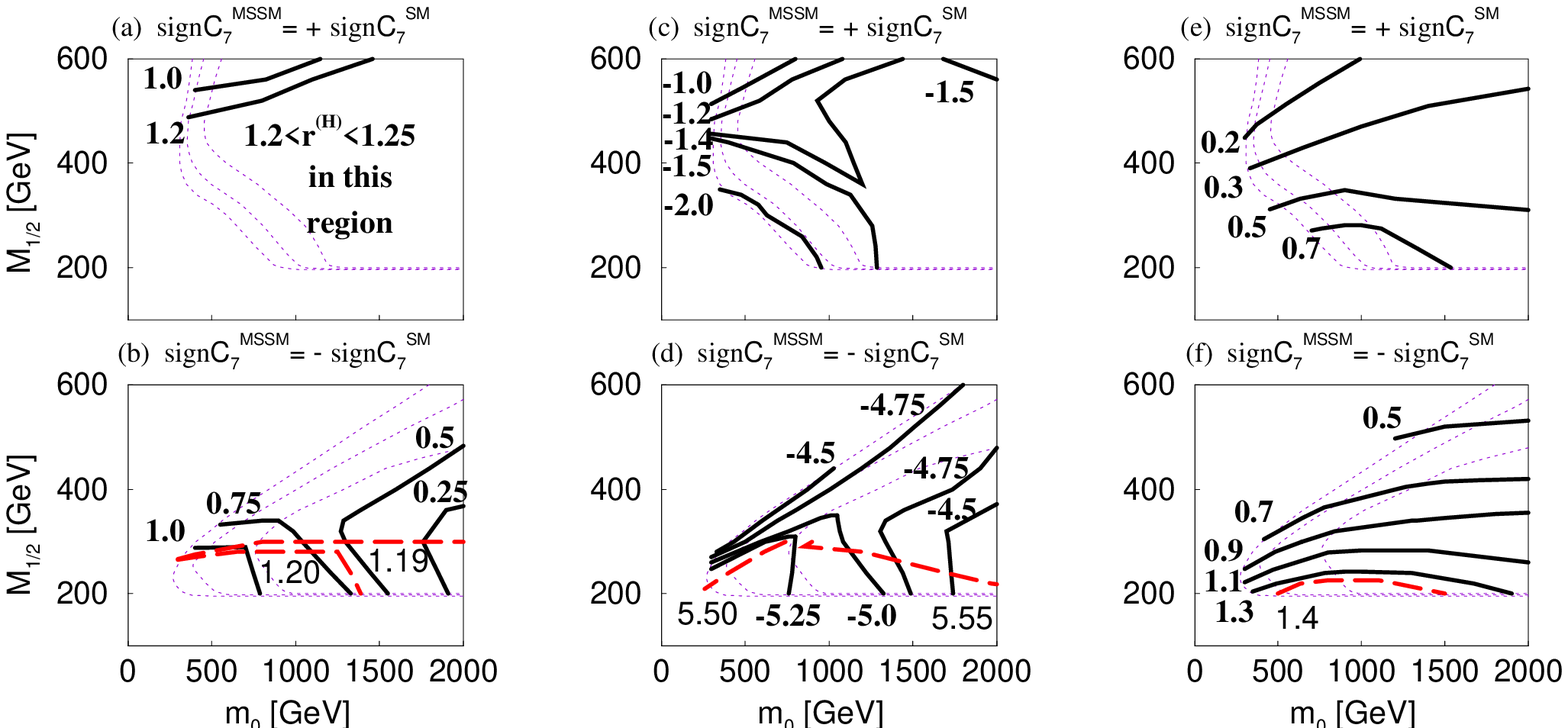
        ,height=3.5cm  
        ,width=10cm   
       }%
}
\caption{ Relative contributions to $C_7$
for the two possible signs of $C_7^{MSSM}$.
Contour plots of 
(a),(b) $r^{(H)}$; 
(c),(d) $r^{(C)}$; 
(e),(f) $r^{(C23)}$;
        with the $\chi^2$ contour lines in the background.}
\label{fig3}
\vskip 11mm
\mbox{%
\epsfig{file=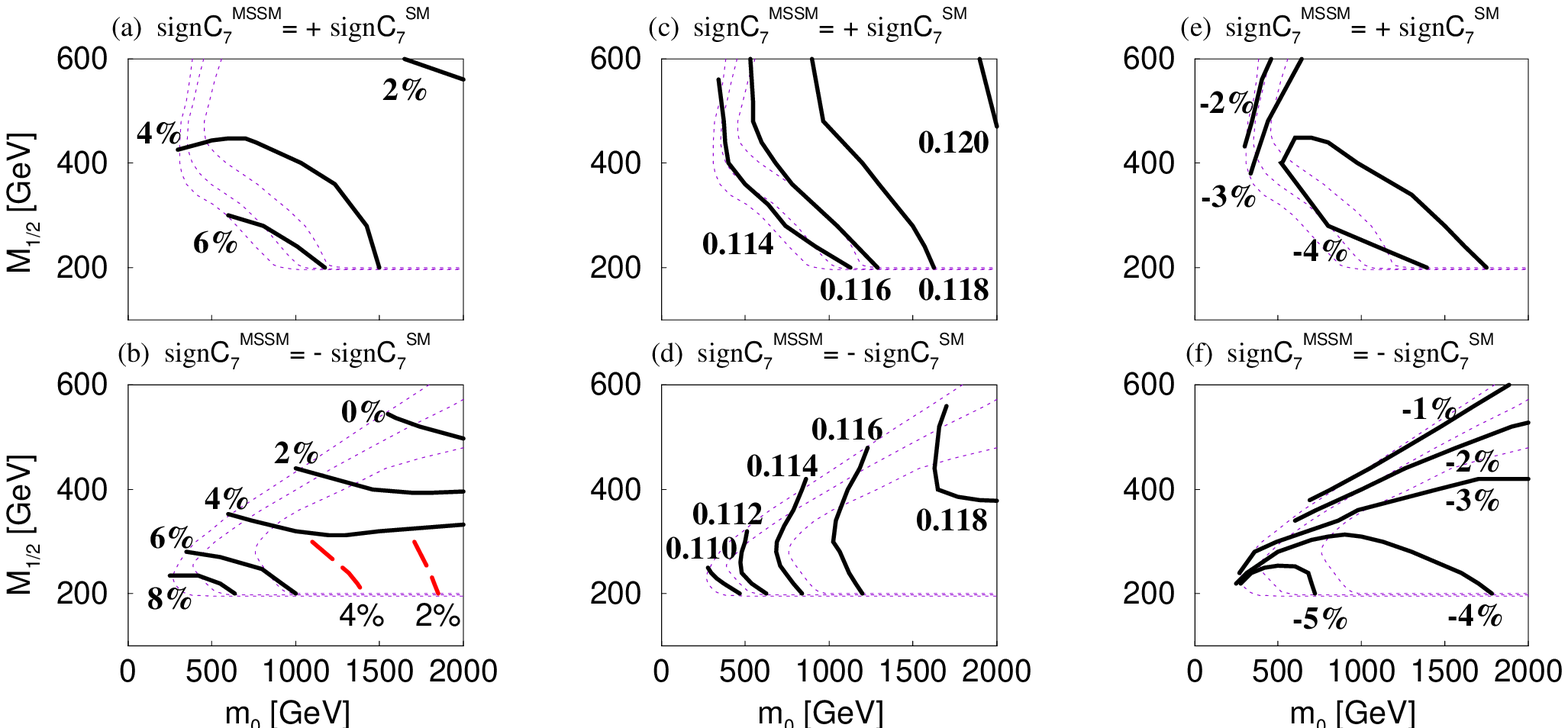
        ,height=3.5cm  
        ,width=10cm   
       }%
}
\caption{ 
Contour plots of 
(a),(b) $\delta m_b^{SUSY}$; 
(c),(d) $\as(M_Z)$; 
(e),(f) GUT threshold $\epsilon_3$;
        with the $\chi^2$ contour lines in the background.}
\label{fig4}
\end{figure}

These expectations are indeed realized in the best fits of model 4c.
Figure 2a (2b) corresponds to the case when the $\bsgam$ amplitude in MSSM
is of the same (opposite) sign as the SM amplitude. As anticipated, these 
two cases work in complementary regions of parameter space.
Figures 2c-d show how well the fits describe 
the measured value of the $BR(\bsgam)$. 
In these figures (and similarly in the
following ones) we show the $\chi^2$ contour lines of fig's 2a-b in the 
background for reference. 
As can be seen, the $BR(\bsgam)$ presents a major constraint since 
whenever the $\chi^2$ values go up, the agreement with the observed $\bsgam$ 
decay rate gets worse. 
For the flipped sign of the amplitude it is interesting that the lightest 
stop $\tilde{t}_1$ can be sufficiently light (see fig.2f) to maintain the 
required large size 
of the chargino contribution even for $m_0$ at $2$TeV. That can be 
arranged due to significant left-right stop mixing and the freedom in varying 
the trilinear scalar coupling $A_0(M_G)$. The corresponding 
$\tilde{t}_L$-$\tilde{t}_R$ mixing angle
changes between 31$^o$ to 40$^o$ across the $(m_0,M_{1/2})$ plane while 
the trilinear coupling $A_t(M_Z)$ ranges from $-300$ to $-1700$GeV
in this case. 
Figure 3 shows the contour plots of constant $r^{(H)}$ and $r^{(C)}$ 
for different signs of $C_7^{MSSM}$. One can compare the numerical results in 
figures 3a {\em vs.} 3c and 3b {\em vs.} 3d 
with the approximate relations (\ref{r_C_ac}) and (\ref{r_C_AC}).

There is one striking feature which is common to both cases. It is that both 
fits would like to have the $BR(\bsgam)$ below rather than above the CLEO
experimental value $2.32\times 10^{-4}$. \cite{bsgam_exp_CLEO} 

In the first case, the tan$\beta$
enhanced chargino contribution tends to be too large when going against 
the charged Higgs contribution, since the latter is not tan$\beta$ enhanced.
The fit clearly tries to make the Higgs contribution as large as possible 
(see fig.3a). As a result, the charged Higgs 
(and then also the whole Higgs sector) tends to be light. 
We get, for instance, the best fit value of the pseudoscalar 
mass $m_A<100$GeV everywhere in the $(m_0,M_{1/2})$ plane.  

In the second case, when the sign of $C_7$ is flipped by the chargino 
contribution, this contribution tends to be not big enough, especially 
if $M_{1/2}>300$GeV. Hence $\bsgam$ would prefer a heavy charged Higgs. 
However, one cannot have good fits with the Higgs sector much heavier than 
squarks$\,$\cite{gaGUT} and thus $r^{(H)}$ varies quite a bit across
the SUSY parameter space (see fig.3b). When $M_{1/2}$ gets below 
$300$GeV, $\bsgam$ is no longer a strong constraint and two separate 
minima can be found in the course of optimization. The two fits work
equally well ($\chi^2$ in each case stays below 1 per 3 d.o.f.) and  
differ only by the 
best fit values of the Higgs masses. One minimum corresponds to $m_{h^0}$ 
and $m_{A}$ settled at the experimental lower limit 
($65$GeV, in our analysis) 
while these masses gradually rise up to $700$GeV in the second ``valley''. 
In the allowed corner with $m_0<700$GeV, the two ``valleys'' approach each 
other and eventually coincide.
The effect of the doubled minimum is indicated in figures 
3b, 3d, 3f and 4b with the solid black (dashed gray) contour lines 
corresponding to the heavier (lighter) Higgs sector.

Can the NLO QCD corrections to BR($\bsgam$) improve the fits?
The calculation of these corrections has not been completed 
in the MSSM. Here we just summarize\cite{bsgamPL} that the rate goes up if
the unknown MSSM matching corrections satisfy 
$C_k^{(MSSM)\,(1)}/\,C_k^{(SM)\,(1)} \stackrel{>}{\sim} -1$, ($k=7,8$), in
the case when \sign$\,C_7^{(MSSM)} =  +$ \sign$\,C_7^{(SM)}$.
Thus the allowed SUSY parameter space is likely to increase in this case.
In the second case, with the opposite sign of the $\bsgam$ amplitude,
the NLO correction will more likely make the fit worse. In this case, 
the net NLO correction increases the rate only if
$C_k^{(MSSM)\,(1)}/C_k^{(SM)\,(1)} \stackrel{<}{\sim} -7$, $k=7,8$. 

In summary, if the future experimental analysis confirms the discrepancy
between the CLEO measured value and the NLO SM calculation \cite{NLO_SM},
the MSSM with large tan$\beta$ could be the solution. Similarly, if
the NLO calculation is completed for the MSSM, and if it turns out to
enhance the LO result as occurred for the SM, then the large tan$\beta$ 
regime will apparently have no problem fitting the $\bsgam$ rate exactly. 

\subsection{Role of $\tilde{c}$-$\tilde{t}$ mixing for $BR({\bsgam})$}
In analogy to previously defined $r^{(H)}$ and $r^{(C)}$ we introduce
$r^{(C23)}\! = C_7^{(C23)} \, / \:  C_7^{(SM)}$, 
where $C_7^{(C23)}$ is the $\tilde{c}_L$-$\tilde{t}_L$ mixing contribution 
to the coefficient $C_7$ at the scale $M_Z$. We show the contour plots 
of the constant 
$r^{(C23)}$ in fig's 3e-f. While the dominant 
chargino contribution is that proportional to the 
$\tilde{t}_L$-$\tilde{t}_R$ mixing, 
$C_7^{(C23)}$ becomes important because of the destructive 
character of the interference among the partial amplitudes. As one can see,
the $\tilde{c}_L$-$\tilde{t}_L$ mixing term can be comparable with 
the SM and $H^-$ contributions. More importantly, it always interferes 
constructively with them. In the case when 
\sign$\,C_7^{(MSSM)} =  +$ \sign$\,C_7^{(SM)}$, this term
helps to counterbalance the large contribution of the left-right stop
mixing. In the complementary case,
when the chargino contribution flips the sign
of the net amplitude the $\tilde{c}_L$-$\tilde{t}_L$ mixing makes
it more difficult to happen. As a result, it has different consequences for
the two fits in figures 2a and 2b, especially important in the region 
($m_0\leq1000$GeV,$\,M_{1/2}\approx 350$GeV) 
where the fits start getting worse. 
These observations are, however, model dependent and their validity relies
on the boundary conditions assumed at the GUT scale.

\subsection{Correlations among $\epsilon_3$, $\as(M_Z)$ and 
            $\delta m_b^{SUSY}$}
\label{ss_e3}

The results on $\bsgam$ show that it is important to have a destructive 
interference among the partial contributions to $\bsgam$. 
However, with the universal boundary
conditions at the GUT scale that can be arranged only for a
specific sign of the $\mu$ parameter: $\mu>0$ in our conventions.
That in turn 
correlates with the positive sign of the SUSY corrections to the $b$
quark mass. This fact implies strong constraints on the SUSY parameter space 
from $m_b$ since $\delta m_b^{SUSY}$ gets a dominant contribution 
from the tan$\beta$ enhanced gluino exchange.
The gluino correction 
can be explicitely suppressed by heavy squarks and by the 
chargino correction which enters with the opposite sign.
It can also be reduced by lower values of $\as(M_Z)$ because of 
the strong coupling in the vertices of the gluino -- $b$-squark diagram.
The contour plots of the net SUSY correction to $m_b(M_Z)$ in the best
fits are shown in figures 4a-b where one can see that these 
effects are quite effective in reducing $\delta m_b^{SUSY}$.

Note that the effect of large positive 
$\delta m_b^{SUSY}$ is also reduced by a lower value 
of $\as(M_Z)$ indirectly ---  due to the RG evolution of the current $b$ mass 
from $M_Z$ down to $M_b$. 
In our analysis, which assumes larger uncertainty for $\as$ than for
$m_b(M_b)$, it pushes $\as(M_Z)$ down (see figures
4c-d). One can trade lower values of $\as$ 
for higher values of $m_b$ provided a smaller uncertainty $\sigma(\as)$
is assumed.
Rather low values of $\as(M_Z)$ are, however, difficult to obtain from
an exact gauge coupling unification. Our analysis shows that 
a few per cent negative correction to $\as(M_G)$ is enough to
get $\as(M_Z)\stackrel{<}{\sim}0.118$. 
The best fit values of $\epsilon_3$ are presented 
in figures 4e-f. Clearly, as squarks get lighter
the SUSY correction to $m_b(M_Z)$ (fig's 4a-b) has to be increasingly 
reduced by the lower values of $\as(M_Z)$ (fig's 4c-d), which in 
turn requires a more substantial departure from the gauge coupling unification 
(fig's 4e-f). $\epsilon_3$ can be 
generated by the spread in masses of heavy states integrated out at
the GUT scale \cite{lucas}. 

\section{Conclusions}
At the present time, when direct evidence for physics beyond the SM
evades experimental observations, global analysis serves as the best
test of new physics. Our results show that minimal SUSY SO(10) models
remain among the candidates for the theory below the Planck scale
and that they can provide specific guidelines in the search 
for signatures of new physics at the electroweak scale.

\section{Acknowledgment}
T.B. would like to thank the organizers for the invitation to a nice
and well organized workshop at the Universita Autonoma de Barcelona.
Thanks goes also to Marcela Carena and Carlos Wagner who collaborated
on parts of this project. This work was supported in part by the U.S.
Department of Energy under contract numbers DE-FG02-91ER40661 and
DOE/ER/01545-728.

\end{document}